\documentstyle[pre,aps,multicol,epsfig]{revtex}
\begin{document}
\draft

\title{Transport Theory in the Context of the \\Normalized Generalized 
Statistics}
\author{F. Q. Potiguar and U. M. S. Costa}
\address{Universidade Federal do Cear\'a, Departamento de F\'\i sica,
Campus do Pici, 60455-760 Fortaleza, Ce, Brazil }
\date{\today}
\maketitle

\begin{abstract}
In this work assuming valid the
equipartition theorem and using the normalized q-expectation 
value, we obtain, until first order approximation, the hydrodynamics
equation for the generalized statistics. This equations are different
from those obtained in the context of the Boltzmann-Gibbs statistics. 
This difference is that now appears two transport coefficient that
depend on the q-value. 
\end{abstract}
\begin{multicols}{2}
\narrowtext
\section{Introduction}
The Generalized Statistics was first proposed by Tsallis \cite{Tsallis} in 1988. It has attracted a lot of attention lately for it has been well applied to many systems, like fully developed turbulence \cite{Beck} and self-graviting systems \cite{Plastinos}. For a periodically updated list please see reference \cite{Lista}.\\
This generalization is based on the new entropic form, namely:
\begin{equation}
S_q=k_B\frac{1-\sum_{i=1}^{W}{p_i}^q}{1-q}
\end{equation}
where $k_B$ is the Boltzmann constant, $p_i$ is the probability that the
system is in a state labeled by $i$, $q$ is a real parameter
and the sum runs over all the allowed states. 
This entropy has the essential feature that it is non-extensive (non-additive). 
For two independent systems,i.e., for two systems which have independent 
probabilities: $p^{(AUB)}=p^{(A)}p^{(B)}$, the entropy of the whole 
system is written as:
\begin{equation}
S_q^{(AUB)}=S_q^{(A)}+S_q^{(B)}+(1-q)S_q^{(A)}S_q^{(B)}
\end{equation}
At that time it was used  to calculate the physical quantities average
values without proper normalization.
Later Tsallis, Mendes and Plastino \cite{TMP} have introduced a new form 
of calculating expectation values, the so called normalized q-expectation value.  Let $O$ be some physical observable, its expectation value, in this context, is given by:
\begin{equation}
<O>_q=\frac{\sum_{i=1}^{W}{p_i}^qO_i}{\sum_{i=1}^{W}{p_i}^q}
\end{equation}
where:
\begin{equation}
p_i=\left[1-(1-q)\beta (H_i-U_q)\right]^{\frac{1}{1-q}}
\end{equation}
is the canonical ensemble distribution function, with $U_q$ being the 
q-average internal energy known, a priori, from experiment.\\
Tsallis introduced a cut-off condition \cite{Tsallis} in this formalism. 
It says that if the argument of the distribution function is negative, 
this distribution vanishes, in other words, if $1-(1-q)\beta (H_i-U_q)<0$, 
$p_i=0$.\\
In 1999, Boghosian \cite{Boghosian} has obtained the hydrodynamic equations 
for the generalized statistics. He calculated them using the unnormalized 
q-expectation value and the well known Chapman-Enskog expansion. He showed 
that the equation of state and the internal energy for the ideal gas 
are q-dependent. He showed also that, in first order approximation, 
the conservation equations don't change and, in second order approximation, 
only the energy conservation equation change.\\
The first change is that the thermal conductivity coefficient is now q-dependent.
It is given by, in the three-dimensional case:
\begin{equation}
K_B=\frac{1}{A_q+(1-q)}\frac{5\tau\rho}{3m\beta}
\end{equation}
where:
\begin{equation}
A_q=1+(1-q)\frac{3}{2}
\end{equation}
$\tau$ is the expansion parameter, $\rho$ is the mass density of the 
system and $\beta$ is the inverse of the temperature.\\
There was, also, the appereance of a new term in this equation, a gradient of 
pressure, with a new transport coefficient, which was called anomalous transport
coefficient:
\begin{equation}
T_q=\frac{1-q}{A_q[A_q+(1-q)]}\frac{5\tau\rho}{3m\beta}
\end{equation}
Obviously this coefficient vanishes in the extensive limit $q\rightarrow1$.\\
We propose in this paper to obtain the hydrodynamic equations using a 
different method and using the normalized q-expectation value. 
We'll see that these new features will led us to the correct equation 
of state for the ideal gas and to different dependence 
on $q$ of the transport coefficient. In section two we obtain the
Conservation Theorems for the conserved quantities in molecular
collisions, we make the zero order approximation,
that corresponds to the first order approximation in
Boghosian\cite{Boghosian} calculations and we present the first order 
approximation (Boghosian second order approximation).
Finally we presents our results and conclusion in section three.\\

\section{Normalized Hydrodynamic Equations}
\subsection{Conservation Theorems}
Here we present our way to obtain the hydrodynamic equations. 
We'll follow reference \cite{Huang} and assume the energy equipartition theorem
to be valid \cite{Martinez}. We are studying an ideal gas in the 6-dimensional
phase space. The one particle hamiltonian for this system reads: 
\[
H=\frac{m}{2}U^2
\]
where $U_i=v_i-u_i$, with $u_i=<v_i>$ the average velocity and $<U_i>=0$, $m$
stands for the mass of the molecule and the index $i$ runs from 1 to 3 and
indicates the components of the velocity.\\ 
It consists in obtaining the equations
which give the transport of the molecular quantities that are conserved in a
molecular collision. To do this, we assume first that we are dealing only with
binary collision and that the two colliding molecules are labeled $1$ and $2$.
We are studying only the molecule $1$. We also assume that the molecular chaos
is still valid.\\
To obtain the conservation theorems we consider the Boltzmann transport
equation:
$$
\left(\frac{\partial}{\partial t}+v_i\partial _i+\frac{F_i}{m}
\partial _{v_i}\right)F_1 =
$$
\begin{equation}
\int d^3v_2d^3v_{f_1}d^3v_{f_2}
\delta ^4(V_f-V_i)|T_{fi}|^2\left(F_{f_2}F_{f_1}-F_2F_1\right)
\end{equation}
where $v_i$ are the components of the thermal velocity of the molecules, 
$F_i$ are the components of the external forces, which are considered 
here velocity independent, the index $2, f_1,f_2$ refers to the second 
molecule before the collision and to the both molecules after the collision, 
the 4-dimensional delta function and the transition matrix elements $T_{fi}$ are related to the 
differential scattering cross section of the collision and the term between 
parenthesis is the molecular chaos assumption. $F_1$ is the particle density
which is the solution for (8) and is unknown.\\
The equilibrium distribution function is given by the escort distribution:
\begin{equation}
F_q=n\frac{\left[1-\frac{1-q}{A_q}\beta H\right]^{\frac{q}{1-q}}}
{\int_{-\infty}^{+\infty}\left[1-\frac{1-q}
{A_q}\beta H\right]^{\frac{q}{1-q}}d^3U}
\end{equation}
where $\beta$ is the inverse of the temperature $\beta =\frac{1}{k_B\beta}$.
The factor $n$ is the particle density. It appears here because 
this distribution is described in the one particle phase space, 
so it is not simply a probability distribution, it is the number 
of particles in the system which have position between $\vec x$ 
and $\vec x+d\vec x$ and velocity between $\vec v$ and $\vec v+d\vec v$. 
This function is related to the  escort probability distribution through 
the relation $F_q=nP_q$. The normalization condition reads $\int F_qd^3U=n$.\\
We define a collision conserved quantity $\chi (\vec x, \vec v,t)$, which have 
the conservation property $\chi_1 +\chi_2=\chi _{f_1}+\chi _{f_2}$. 
The following theorem can be proved:
\begin{equation}
\int d^3v_1\chi _1(\vec x,\vec v)\left(\frac{\partial F_1}{\partial t}
\right)_{coll}=0
\end{equation}
where $\left(\frac{\partial F_1}{\partial t}\right)_{coll}$ is the 
collisional integral, making the changing of variables below:\\
First: $\vec v_1 \rightarrow \vec v_2$ and vice-versa.\\
Second: $\vec v_1 \rightarrow \vec v_{f_1}$, $\vec v_2\rightarrow\vec v_{f_2}$ 
and vice-versa.\\
Third: $\vec v_1\rightarrow\vec v_{f_2}$, $\vec v_2\rightarrow\vec v_{f_1}$ 
and vice-versa.\\
As these changes don't affect the transition matrix elements $T_{fi}$ and 
the delta function, we add each of the integrals obtained with equation (10)and divide the 
result by 4. We end up with:
\[\int d^3v\chi\left(\frac{\partial F_1}{\partial t}\right)_{col}=\frac{1}{4}
\int d^3v_1d^3v_2d^3v_{f_1}d^3v_{f_2}\delta ^4(P_f-P_i)\]
\[\times|T_{fi}|^2\left(F_{f_2}F_{f_1}-F_2F_1\right)(\chi_1 +\chi _2
-\chi _{f_1}-\chi _{f_2})\]
Through the conservation of $\chi_1$, the theorem is proved.\\
Using equation (8), we have, changing the volume element form $d^3v$ to $d^3U$:
\begin{equation}
\int d^3U_1\chi_1\left(\frac{\partial}{\partial t}+v_i\partial _i+
\frac{F_i}{m}\partial _{v_i}\right)F_1(\vec x,\vec v,t)=0
\end{equation}
Defining the expectation value as:
\begin{equation}
<O>=\frac{1}{n}\int F_1Od^3U
\end{equation}
and rearranging terms in (11), we have the desired conservation theorem:
$$\frac{\partial}{\partial t}<n\chi_1 >+\partial _i<nv_i\chi_1 >-n<v_i
\partial _i\chi_1 >$$
\begin{equation}
-\frac{n}{m}F_i<\partial _{v_i}\chi_1 >=0
\end{equation}
Now we calculate the above equation for the three conserved quantities, 
namely mass, momentum and energy.\\
For the mass, $\chi_1 =m$, we have:
\begin{equation}
\frac{\partial \rho}{\partial t}+\partial _i(\rho u_i)=0
\end{equation}
where:
\begin{equation}
\rho =mn(\vec x,t)
\end{equation}
is the mass density of the system.\\
For the momentum,  $\chi_1 =mv_i$, we get:
\begin{equation}
\frac{\partial}{\partial t}(\rho u_i)+\partial _j(\rho <U_iU_j>)
+\partial _j (\rho u_iu_j)=\rho \frac{F_i}{m}
\end{equation}
Finally for the energy,  $\chi _1 =\frac{m}{2}U^2$, we have, 
using the mass conservation equation (14):
\[\frac{\partial}{\partial t}
\left(\frac{1}{\beta}\right)
+u_i\partial _i\left(\frac{1}{\beta}\right)\]
\begin{equation}
+\frac{2m\rho}{3}\partial _iu_j<U_iU_j>+\frac{m}{3}\partial _i(\rho <U_iU^2>)=0
\end{equation}
where we used the fact that the internal energy for this system is given by 
the classical energy equipartition theorem.
All we have to do now is calculate the two expectation values in order 
to obtain the hydrodynamic equation for the generalized statistics.

\subsection{The Zero Order Approximation}
To obtain the exact equations, we should use the solution for (8) $F_1$. As we
don't still have this solution, we must approximate it. We assume that the
system, locally, obeys equation (9). This is the local equilibrium assumption. We impose a space and time dependence of 
the macroscopic parameters number density $n$, average velocity $u_i$ 
and temperature $\frac{1}{\beta}$. Hence the distribution function reads:
\[{F_q}^{(0)}=n\frac{{\left[f^{(0)}\right]}^{\frac{q}{1-q}}}
{\int {\left[f^{(0)}\right]}^{\frac{q}{1-q}}d^3U}=
\frac{n}{I}{\left[f^{(0)}\right]}^{\frac{q}{1-q}}\]
The integral in the denominator is readily calculated:
\[I=\int \left[1-\frac{1-q}{A_q}\beta \frac{m}{2}U^2\right]^{\frac{q}{1-q}}d^3U\]
Using the spherical volume element and making the following change of variable:
\begin{equation}
U=\left(\left|\frac{A_q}{1-q}\right|\frac{2}{m\beta}\right)^{1/2}x^{1/2}
\end{equation}
we have:
\begin{equation}
I=2\pi\left(\left|\frac{A_q}{1-q}\right|\frac{2}{m\beta}\right)^{3/2}\int x^{1/2}(1-\sigma x)^{\frac{q}{1-q}}dx
\end{equation}
where $\sigma$ is the signal of $\frac{1-q}{A_q}$. It should be noticed that
this normalization factor depends only on the temperature, as in BG statistics.\\
We must calculate the quantity $<U_iU_j>_q$:
\[<U_iU_j>_q=\frac{1}{n}\int U_iU_j{F_q}^{(0)}d^3U\]
\[=\frac{1}{I}\int_{-\infty}
^{+\infty}U_iU_j\left[1-\frac{1-q}{A_q}\beta \frac{m}{2}U^2\right]
^{\frac{q}{1-q}}d^3U\]
Utilizing parity arguments we can show that this integral will be non-zero 
only for $i=j$, so we have, changing volume element and the variables 
as was done in the last calculation we obtain:
\[<U_iU_j>_q=\frac{\delta _{ij}}{3}\left|\frac{A_q}{1-q}\right|
\frac{2}{m\beta}\frac{\int x^{3/2}(1-\sigma x)^{\frac{q}{1-q}}dx}
{\int x^{1/2}(1-\sigma x)^{\frac{q}{1-q}}dx}\]
For $\sigma =+1$, these integrals are made in the interval between 0 and 1, 
due to Tsallis cut-off condition, and give the result:
\begin{equation}
<U_iU_j>_q=\frac{\delta _{ij}}{m\beta}
\end{equation}
For $\sigma =-1$, the integrals are calculated between 0 and $+\infty$ 
and give the same result.\\
The other expectation value, $<U_iU^2>_q$ vanishes in this approximation 
because it is an odd function.\\
The mass conservation equation is the same in all orders of approximation:
\[\frac{\partial \rho}{\partial t}+\partial _i(\rho u_i)=0\]
The momentum conservation equation is written as:
\[\frac{\partial}{\partial t}(\rho u_i)+\partial _j(\rho u_iu_j)+\partial _j
\left(\delta _{ij}\frac{\rho}{m\beta}\right)=\rho\frac{F_i}{m}\]
Using the mass equation, we have:
\begin{equation}
\left[\frac{\partial}{\partial t}+u_j\partial _j\right]u_i+\frac{1}{\rho}
\partial _i\left(\frac{\rho}{m\beta}\right)=\frac{F_i}{m}
\end{equation}
The term $\frac{\rho}{m\beta}$ is defined as the pressure of the system:
\[P=\frac{n}{\beta}\]
\begin{equation}
PV=Nk_BT
\end{equation}
We recognize this equation as the equation of state of the ideal gas. 
As we can see it is independent of $q$.\\
The energy conservation equation is written as:
\[\rho\frac{\partial}{\partial t}\left(\frac{1}{\beta}\right)+\frac{1}{\beta}
\frac{\partial\rho}{\partial t}+\rho u_i\partial _i\left(\frac{1}{\beta}\right)
+\frac{1}{\beta}\partial _i(\rho u_i)+\frac{2}{3}\delta _{ij}\frac{\rho}{\beta}
\partial _ju_i=0\]
Again using the equation for the mass and dividing by $\rho$:
\begin{equation}
\left[\frac{\partial}{\partial t}+u_i\partial _i\right] \frac{1}{\beta}
+\frac{2}{3}(\partial _iu_i)\frac{1}{\beta}=0
\end{equation}
We conclude that, in the zero order approximation, the hydrodynamic 
equations are q-invariant and we obtained the correct equation of state 
for the ideal gas.

\subsection{The First Order Approximation}
To obtain the fully hydrodynamic equations a solution of Boltzmann 
transport equation is needed to be used as the distribution function. 
Since this equation until now was not exactly solved, 
we made an approximation in the previous section. Now we add to this zero order
approximation a correction to it. This procedure consist in writing this exact function as:
\[F_1={F_q}^{(0)}+{F_q}^{(1)}\]
where ${F_q}^{(1)}$ is the correction to the zero order approximation. 
It can be shown that this correction is given by the following
relation\cite{Huang}:
\[{F_q}^{(1)}=-\tau\left(\frac{\partial}{\partial t}+v_i\partial _i+
\frac{F_i}{m}\partial _{v_i}\right){F_q}^{(0)}\]
where $\tau$ is a parameter of the order of magnitude of the collision time. 
We must calculate each of the derivatives of ${F_q}^{(0)}$. 
At first we evaluate the following derivatives. For the density we have:
\[\frac{\partial {F_q}^{(0)}}{\partial\rho}=\frac{\partial}{\partial\rho}
\left[\frac{\rho}{mI}{f^{(0)}}^{\frac{q}{1-q}}\right]\]
\begin{equation}
\frac{\partial {F_q}^{(0)}}{\partial\rho}=\frac{{F_q}^{(0)}}{\rho}
\end{equation}
Second the average velocity:
\[\frac{\partial {F_q}^{(0)}}{\partial u_i}=\frac{\partial}{\partial u_i}\left[\frac{\rho}{mI}{f^{(0)}}^{\frac{q}{1-q}}\right]\]
\begin{equation}
\frac{\partial {F_q}^{(0)}}{\partial u_i}=\frac{q}{A_q}\beta mU_i\frac{{F_q}^{(0)}}{f^{(0)}}
\end{equation}
Third the temperature:
\[\frac{\partial {F_q}^{(0)}}{\partial\left(\frac{1}{\beta}\right)}=\frac{\partial}{\partial\left(\frac{1}{\beta}\right)}\left[\frac{\rho}{mI}{f^{(0)}}^{\frac{q}{1-q}}\right]\]
\begin{equation}
\frac{\partial {F_q}^{(0)}}{\partial\left(\frac{1}{\beta}\right)}=\frac{q}{A_q}{\beta}^2\frac{m}{2}U^2\frac{{F_q}^{(0)}}{f^{(0)}}-\frac{3}{2}\beta {F_q}^{(0)}
\end{equation}
The derivative of the distribution function with respect to the thermal 
velocity of the molecules is simply the derivative with respect to the 
average velocity with a changed signal:
\begin{equation}
\frac{\partial {F_q}^{(0)}}{\partial v_i}=-\frac{q}{A_q}\beta mU_i
\frac{{F_q}^{(0)}}{f^{(0)}}
\end{equation}
Using these results in the first order approximation function we get:
\begin{eqnarray}
{F_q}^{(1)}=-\frac{\tau}{\rho}\left(1-\frac{q}{A_q}\frac{1}{f^{(0)}}\right)U_i
\partial _i\rho{F_q}^{(0)}\nonumber\\
+\tau\frac{q}{A_q}\frac{1}{f^{(0)}}\beta m
\left(U_iU_j-\frac{\delta _{ij}}{3}U^2\right)\partial _iu_j{F_q}^{(0)}
\nonumber\\
+\tau\beta\left[\frac{q}{A_q}\frac{1}{f^{(0)}}\left(\beta\frac{m}{2}U^2-1\right)
-\frac{3}{2}\right]U_i\partial _i\frac{1}{\beta}{F_q}^{(0)}
\end{eqnarray}
This function is different from the one calculated by the 
Maxwell-Boltzmann function \cite{Huang}, it has the explicit dependence 
on $q$ through the term $\frac{q}{A_q}\frac{1}{f^{(0)}}$, which is unity 
in the extensive limit.\\
The quantity  $<U_iU_j>_q$, in this order of approximation, is given by:
\[<U_iU_j>_q=\frac{1}{n}\left[\int U_iU_j{F_q}^{(0)}d^3U+\int U_iU_j
{F_q}^{(1)}d^3U\right]\]
The first term is simply $\frac{\delta _{ij}}{m\beta}$. The second term 
will give non-vanishing contribution only through the second term from 
equation (28), the others are odd functions:
\[<U_iU_j>_q=-\frac{q}{A_q}\tau\beta \frac{m}{n}\partial _lu_k
\left[\int U_iU_jU_kU_l\frac{{F_q}^{(0)}}{f^{(0)}}d^3U- \right. \]
\[\left.\frac{\delta _{kl}}{3}\int U_iU_jU^2\frac{{F_q}^{(0)}}{f^{(0)}}d^3U\right]\]
This equation is a traceless tensor and it is suffices to calculate any 
component of it. Putting $i=1$ and $j=2$, we have the result:
\[<U_1U_2>_q=-\frac{q}{A_q}\tau\beta \frac{m}{I}\partial _lu_k
\int U_1U_2U_kU_l\]
\[\times \left[1-\frac{1-q}{A_q}\beta\frac{m}{2}U^2\right]^
{\frac{q}{1-q}-1}d^3U\]
This integral is non-zero only for $k=1,2$ and $l=1,2$. We have, writing 
$U_1=Usen\theta cos\phi$, $U_2=Usen\theta sen\phi$ and changing the volume 
element to a spherical one:
\[<U_1U_2>_q=-\frac{q}{A_q}\tau\beta \frac{m}{I}(\partial _1u_2+\partial
_2u_1)\]
\[\times\int U^6
\left[1-\frac{1-q}{A_q}\beta\frac{m}{2}U^2\right]^{\frac{q}{1-q}-1}dU\]
\[\times\int_{0}^{\pi}{sen}^5\theta d\theta\int_{0}^{2\pi}sen^2\phi 
cos^2\phi d\phi\]
Using equations (18) and (19), we have:
\[<U_1U_2>_q=-\frac{q}{A_q}\tau\beta \frac{m}{15}
(\partial _1u_2+\partial _2u_1)
\left(\frac{A_q}{1-q}\right)^2\]
\[\times\left(\frac{2}{m\beta}\right)^2
\frac{\int x^{5/2}(1-\sigma x)^{\frac{q}{1-q}-1}dx}
{\int x^{1/2}(1-\sigma x)^{\frac{q}{1-q}}dx}\]
For the $\sigma =+1$ case, the above integrals give:
\[<U_1U_2>_q=-\frac{\tau}{m\beta}(\partial _1u_2+\partial _2u_1)\]
For the other case, $\sigma =-1$:
\[<U_1U_2>_q=-\frac{\tau}{m\beta}(\partial _1u_2+\partial _2u_1)\]
So the value of $<U_iU_j>_q$ in this order can be written as:
\begin{equation}
<U_iU_j>_q=\frac{\delta _{ij}}{m\beta}-\frac{\tau}{m\beta}\left(\partial _iu_j+
\partial _ju_i-\frac{2}{3}\delta _{ij}\partial _ku_k\right)
\end{equation}
This term doesn't change in this context.\\
The momentum conservation energy is written as:
$$
\frac{\partial}{\partial t}(\rho u_i)+\partial _j\left[\frac{\rho}{m\beta}-
\frac{\tau\rho}{m\beta}\left(\partial _iu_j+\partial _ju_i\right.\right.$$
\begin{equation}
\left.\left.-\frac{2}{3}
\delta _{ij}\partial _ku_k\right)\right]+\partial _j(\rho u_iu_j)
=\rho\frac{F_i}{m}
\end{equation}
where the term $\frac{\tau\rho}{m\beta}$ is the viscosity coefficient 
of this system:
\begin{equation}
\mu=\frac{\tau\rho}{m\beta}
\end{equation}
The other expectation value to be calculated is given by, where only the 
first and third terms of (28) will give non-vanishing contributions:
\[<U_iU^2>_q=-\frac{\tau}{n}\frac{\delta _{ij}}{3}\left(\int U^4{P_q}^
{(0)}d^3U\right.\]
\[\left.-\frac{q}{A_q}\int U^4\frac{{P_q}^{(0)}}{f^{(0)}}d^3U\right)
+\beta\left(\beta\frac{m}{2}\frac{q}{A_q}\int U^6\frac{{P_q}^{(0)}}
{f^{(0)}}d^3U\right.\]
\[\left.-\frac{q}{A_q}\int U^4\frac{{P_q}^{(0)}}{f^{(0)}}d^3U-
\frac{3}{2}\int U^4{P_q}^{(0)}d^3U\right)\partial _j\frac{1}{\beta}\]
Calculating each integral separately, we have:
\[I_1=\int U^4{F_q}^{(0)}d^3U\]
This integral is calculated in the same way that the others above, 
changing the volume elements, using equation (17) and observing 
the cut-off condition, we obtain:
\begin{equation}
I_1=n\frac{A_q}{A_q+(1-q)}\frac{15}{{(m\beta)}^2}
\end{equation}
The second integral is given by:
\[I_2=\frac{q}{A_q}\int U^4\frac{{F_q}^{(0)}}{f^{(0)}}d^3U\]
which results in:
\begin{equation}
I_2=n\frac{15}{{(m\beta)}^2}
\end{equation}
The last one is given by:
\[I_3=\beta\frac{m}{2}\frac{q}{A_q}\int U^6\frac{{F_q}^{(0)}}{f^{(0)}}d^3U\]
Resulting in:
\begin{equation}
I_3=n\frac{105}{2{(m\beta)}^2}\frac{A_q}{A_q+(1-q)}
\end{equation}
Using the above results, the expectation value $<U_iU^2>_q$ is given by:
$$<U_iU^2>_q=\frac{1-q}{A_q+(1-q)}\frac{5\tau}{\rho{(m\beta)}^2}
\partial _i\rho$$
\begin{equation}
-\frac{A_q-(1-q)}{A_q+(1-q)}\frac{5\tau}{m^2\beta}
\partial _i\frac{1}{\beta}
\end{equation}
The energy conservation equation is given by:
\[\frac{3}{2m}\frac{\partial}{\partial t}\left(\frac{\rho}
{\beta}\right)+\frac{3}{2m}\partial _i\left(\frac{\rho}{\beta} u_i\right)
+\partial _ju_i\left[\frac{\rho}{m\beta}\right.\]
\[\left.-\frac{\tau\rho}{m\beta}
\left(\partial _iu_j+\partial _ju_i-\frac{2}{3}\delta _{ij}\partial _ku_k\right)\right]+\]
\[+\frac{1}{2}\partial _i\left[\frac{1-q}{A_q+(1-q)}
\frac{5\tau}{{(m\beta)}^2}\partial _i\rho \right.\]
\[\left.-\frac{A_q-(1-q)}{A_q+(1-q)}
\frac{5\tau\rho}{m^2\beta}\partial _i\frac{1}{\beta}\right]=0\]
Using the equation for the mass, we have:
\begin{eqnarray}
\frac{\partial}{\partial t}\frac{1}{\beta}+u_i\partial _i\frac{1}{\beta}
=-\frac{2}{3\beta}\partial _ju_i+\frac{2\tau}{3\beta}\left(\partial _iu_j
+\partial _ju_i \right.\nonumber\\
\left. -\frac{2}{3}\delta _{ij}\partial _ku_k\right)\partial _ju_i
-\frac{1}{\rho}\partial _i\left(K_q\partial _i\frac{1}{\beta}-\frac{T_q}{\rho}
\partial_i\frac{\rho}{\beta}\right)
\end{eqnarray}
where:
\begin{equation}
K_q=\frac{A_q}{A_q+(1-q)}\frac{5\tau\rho}{3m\beta}
\end{equation}
is the q-dependent thermal conductivity coefficient. This coefficient, 
due to its dependence on $q$, can assume positive, $q<1,4$ and $q>1,6$, 
or negative values, $1,4<q<1,6$. These negative values are non physical. 
Also we have defined:
\begin{equation}
T_q=\frac{1-q}{A_q+(1-q)}\frac{5\tau\rho}{3m\beta}
\end{equation}
which is the new transport coefficient. This one too can have positive, $q<1$ e $q>1,4$, as well as negative values, $1<q<1,4$. This expression vanishes, properly, in the extensive limit, $q\rightarrow1$.\\
The relation between these two coefficients is given by:
\begin{equation}
\frac{T_q}{K_q}=\frac{1-q}{A_q}
\end{equation}
The relation between the thermal conductivity and the viscosity coefficients 
is given by:
\begin{equation}
\frac{K_q}{\mu}=\frac{A_q-(1-q)}{A_q+(1-q)}\frac{5}{3}
\end{equation}
So the old relation is now $q$ dependent and is equal to the old value 
in the extensive limit since $A_q\rightarrow1$ when $q\rightarrow1$.\\
The coefficients obtained by Boghosian are related to those obtained 
here through the following relation:
\begin{equation}
\frac{K_q}{K_B}=\frac{T_q}{T_B}=A_q=1+(1-q)\frac{3}{2}
\end{equation}

\section{Conclusions}
The hydrodynamic equations obtained here are similar to those obtained 
by Boghosian \cite{Boghosian}. The same changes in the energy conservation 
equation in first order are observed here, only that we have showed 
that the q-dependence of the thermal conductivity and the new transport 
coefficients are different.We have showed also, through the derivation 
of the conservation equations, 
that the ideal gas state equation is still valid in the generalized 
statistics context.\\ 
As was said there are some differences between our results and the ones
obtained by Boghosian\cite{Boghosian}. The most important is that these
differences in the $q$-dependence of the transport coefficients, in the thermal
conductivity and in the new transport coefficient are due the difference in
the state equation and in the internal energy for the ideal gas. In our result
we do not have intrinsically a $q$-dependence, while in the Boghosian calculation
they have a factor $A_q= 1 +(1-q) \frac{3}{2}$. In order to now what result
is the correct one is necessary to make some application of these transport
equations. In despite of this we believe that our results are better because
we obtain the correct state equation (independent of the parameter $q$) 
and we assume the validity of the classical equipartition theorem.



\acknowledgments
We would like to thank FUNCAP and CNPq ({\it Brazilian Agencies})
 for financial supporting 
and Prof. M. P. Almeida and Evaldo M. F. Curado for enlightening discussions.

\end{multicols}
\end{document}